\documentclass[prl,amsfonts,amsmath,twocolumn,superscriptaddress,nofootinbib]{revtex4}

\usepackage{color}
\usepackage{graphicx} 
\usepackage{epstopdf}%

\usepackage{mathpazo}      

\newcommand {\be} {\begin{equation}} 
\newcommand {\ba}{\begin{eqnarray}} 
\newcommand {\ee} {\end{equation}} 
\newcommand{\ea} {\end{eqnarray}}

\renewcommand{\Im}{{\rm Im\,}} 

\renewcommand{\epsilon}{\varepsilon}

\preprint{MKPH-T-11-03}

\begin{document}

\title{Higher order proton structure corrections to the Lamb shift in muonic hydrogen}

\author{Carl E.\ Carlson}

\affiliation{Helmholtz Institut Mainz, Johannes Gutenberg-Universit\"at, D-55099 Mainz, Germany}
\affiliation{Department of Physics, College of William and Mary, Williamsburg, VA 23187, USA}

\author{Marc Vanderhaeghen}
\affiliation{Institut f\"ur Kernphysik, Johannes Gutenberg-Universit\"at, D-55099 Mainz, Germany}

\date{January 31, 2011}

\begin{abstract}
The recent conundrum with the proton charge radius inspires reconsideration of the corrections that enter into determinations of the proton size.  We study the two-photon proton-structure corrections, with special consideration of the non-pole subtraction term in the dispersion relation, and using fits to modern data to evaluate the energy contributions.  We find that individual contributions change more than the total, and present results with error estimates.
\end{abstract}

\maketitle




The recent measurement of the proton charge radius using the Lamb shift in muonic hydrogen~\cite{Pohl:2010zz} has given a value that is a startling 4\%, or 5 of the previous standard deviations, lower  than the values obtained from energy level shifts in electronic hydrogen~\cite{Mohr:2008fa} or from electron-proton scattering experiments~\cite{Sick:2003gm, Bernauer:2010wm}.  Specifically, the new muonic hydrogen measurement~\cite{Pohl:2010zz} gives 
\be
R_E = 0.84184 \, (67) \, {\rm fm},
\ee
compared to the CODATA value~\cite{Mohr:2008fa}
\be
R_E = 0.8768 \, (69) \, {\rm fm},
\ee
or the latest electron scattering value~\cite{Bernauer:2010wm}
\be
R_E = 0.879 \, (8) \, {\rm\ fm},
\ee
where we have added in quadrature the several uncertainties given in~\cite{Bernauer:2010wm}.

The promise of the muonic hydrogen measurement was that, because a muon would orbit closer to the proton than an electron, the effect of the proton structure on the energy level splittings would be enhanced and a more accurate proton radius could be obtained.  Based on the quoted error limit, that promise has been achieved.  However, the discrepancy from the previous results requires an explanation, and invites a reconsideration of the theoretical corrections that are involved in connecting the experimental energy shift to the proton charge radius~\cite{Jentschura:2010ej,Jentschura:2010ha}.  In this note, we will focus on one of the corrections, namely the order $\alpha^5$ proton size corrections to the Lamb shift.

The leading $\mathcal O(\alpha^4)$ and $\mathcal O(\alpha^5)$ proton structure contributions to the hydrogen Lamb shift are often given as
\be
\Delta E = \frac{2 \pi \alpha}{3} \phi_n(0)^2 \left( R_E^2 
	- \frac{1}{2} m_r \alpha R_{(2)}^3  \right)	,
\ee
where $\phi^2_n(0)$ is the square of the $nS$-state wave function at the origin (which contains a factor $\alpha^3$) and $m_r$ is the lepton-proton reduced mass.  The quadratic term was obtained non-relativistically in~\cite{Karplus:1952zza}, and one can verify from a relativistic calculation that $R_E$ is indeed the proton charge radius~\cite{Eides:2000xc}.

The $\mathcal O(\alpha^5)$ term was given by Friar~\cite{Friar:1978wv} as,
\be
R_{(2)}^3 = \int d^3 r_1 \, d^3 r_2 \,  |\vec r_1 -\vec r_2|^3 \rho_E(r_1) \rho_E(r_2)	.
\ee
where $\rho_E$ is the charge density of the proton itself.  Friar called $R_{(2)}^3$ the third Zemach moment, because it is reminiscent of an integral found by Zemach~\cite{Zemach:1956zz} in the related context of hyperfine splitting.

\begin{figure}[t]
\bigskip
\begin{center}
\includegraphics[width = 60 mm]{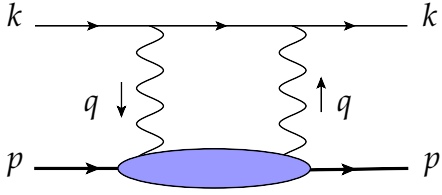}
\caption{The box diagram for the $\mathcal O(\alpha^5)$ corrections.
}
\label{fig:lambbox}
\end{center}
\end{figure}

In modern times, one should calculate the $\mathcal O(\alpha^5)$ corrections field theoretically using the diagram shown in Fig.~\ref{fig:lambbox}, as has been done by Pachucki~\cite{Pachucki:1996zza,Pachucki:1999zza} and by others~\cite{Faustov:1999ga}.  We wish to reexamine the calculation here,  for the purpose of better assessing the connection between the elastic and inelastic contributions, and to better evaluate the subtraction term needed in a dispersion relation that is part of the work.

The calculation of the elastic and inelastic contributions should be done together.  Perhaps in the future a direct QCD calculation will be possible, and there is an exploration of the hadronic corrections to the Lamb shift using chiral perturbation theory~\cite{Nevado:2007dd}, but for the present to obtain the required accuracy the calculation needs to be done dispersively, connecting the off-shell Compton scattering which is the hadronic side of the diagram to information obtained from electron-proton scattering.    In particular, done that way the elastic contributions require no (non-existent) knowledge of form factors for situations where a proton is off-shell.  It also means that certain non-pole contributions to the Compton amplitudes are not picked up by the dispersive calculation and in fact do not contribute.   

We also analyze more concretely the subtraction function that appears because one of the dispersion relations does not converge if unsubtracted.  The subtraction function depends on the photon four-momentum squared, $Q^2$, and its value at $Q^2=0$ is given in terms of the proton magnetic polarizability.  Its $Q^2$ dependence can be estimated by calculating a two-pion loop contribution which couples to the nucleon as a scalar.   One does not need to use a $Q^2$ dependence assumed given by the nucleon electromagnetic form factor, as has been done previously.




The Feynman diagram for the two-photon proton-structure correction to the Lamb shift is shown in Fig.~\ref{fig:lambbox}.  To the level of accuracy needed here, all external lines have zero three-momentum.  The blob corresponds to off-shell forward Compton scattering, given in terms of the Compton tensor
\begin{align}
&T^{\mu\nu}(p,q) = \frac{i}{8\pi M}  \int d^4x	\,e^{iqx}
	\langle p | T j^\mu(x) j^\nu(0) | p \rangle  \nonumber\\[1ex]
	& \quad = \left(-g^{\mu\nu} + \frac{q^\mu q^\nu}{q^2} \right) T_1(\nu,Q^2)
		\nonumber\\
	& \quad + \frac{1}{M^2} \left( p^\mu - \frac{p\cdot q}{q^2} q^\mu \right)
	\left( p^\nu - \frac{p\cdot q}{q^2} q^\nu \right)  T_2(\nu,Q^2)	,
\end{align}
where $q^2 = -Q^2$,  $\nu = p\cdot q/M$, and $M$ is the nucleon mass.   A spin average is implied and the state normalization is 
$\langle p | p' \rangle = (2\pi)^3 \, 2E \, \delta^3 (\vec p- \vec p')$.  The functions 
$T_{1,2}(\nu,q^2)$ are each even in $\nu$ and their imaginary parts are related to the structure functions measured in electron or muon scattering by
\begin{align}
\Im T_1(\nu,Q^2) &= \frac{1}{4M} F_1(\nu,Q^2) ,		\nonumber\\
	\Im T_2(\nu,Q^2) &= \frac{1}{4\nu} F_2(\nu,Q^2) ,
\end{align}
with $\nu > 0$ and where $F_{1,2}$ are standard~\cite{Nakamura:2010zzi}.

After doing a Wick rotation, where $q_0 = i Q_0$ and $\vec Q = \vec q$, one obtains the $\mathcal O(\alpha^5)$ energy shift as
\begin{align}
\Delta E &=  \frac{8\alpha^2 m}{ \pi } \phi^2_n(0)  \int d^4Q 
					\nonumber\\[1ex]
&\times \frac{ (Q^2+2Q_0^2) T_1(iQ_0,Q^2) - (Q^2 -Q_0^2) T_2(iQ_0,Q^2) }
	{ Q^4 (Q^4 + 4m^2 Q_0^2) }	\,,
\end{align}
where $m$ is the lepton mass, and
$
\phi_n^2(0) = {m_r^3 \alpha^3}/{(\pi n^3)}
$
with $m_r = mM/(M+m)$.




\begin{figure}[b]
\begin{center}
\includegraphics[width = 84 mm]{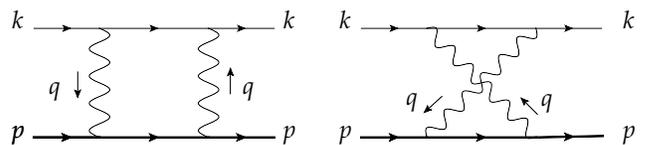}
\caption{Elastic contributions to the box diagram.
}
\label{fig:el}
\end{center}
\end{figure}

The $T_i$ are obtained using dispersion relations.  Regge arguments~\cite{Abarbanel:1967zza} suggest that $T_2$ satisfies an unsubtracted dispersion relation in $\nu$ at fixed $Q^2$, but that $T_1$ will require one subtraction.  Before proceeding, we will note that the Born terms, obtained from the elastic box and crossed box of Fig.~\ref{fig:el} and the vertex function 
$\Gamma^\mu = \gamma^\mu F_1(Q^2) + (i/2M)\sigma^{\mu\nu}q_\nu F_2(Q^2)$ for an incoming photon, are
\begin{align}
T_1^B(q_0,Q^2) &= \frac{1}{4\pi M} \left\{ 
	\frac{Q^4 G_M^2(Q^2) }{(Q^2-i\epsilon)^2 - 4M^2 q_0^2} - F_1^2(Q^2) 
	\right\}		\,,
					\nonumber\\
T_2^B(q_0,Q^2) &= \frac{M  Q^2}{ \pi (1+\tau_p) }  
	\frac{G_E^2(Q^2) + \tau_p G_M^2(Q^2)}
		{(Q^2-i\epsilon)^2 - 4M^2 q_0^2}	\,,
\end{align}
where $\tau_p = Q^2/(4M^2)$, and the electric and magnetic form factors are
\begin{align}
G_E(Q^2) &= F_1(Q^2) - \tau_p F_2(Q^2) , \nonumber\\
G_M(Q^2) &= F_1(Q^2) + F_2(Q^2) .
\end{align}
The Born terms are reliable for obtaining the imaginary parts of the nucleon pole terms, but not reliable in general, since the given vertex assumes the incoming and outgoing nucleons are both on shell.  

Calling the first term in $T_1^B$ the pole term, one can split the whole of $T_1$ into pole term and non-pole terms,
\be
T_1(q_0,Q^2) = T_1^{pole} + \overline T_1  \,.
\ee
The pole term alone evidently allows an unsubtracted dispersion relation, and this term calculated from the dispersion relation simply reproduces itself.  With a once subtracted dispersion relation for $\overline T_1$, one has
\begin{align}
T_1(q_0,Q^2) &= T_1^{pole}(q_0,Q^2) + \overline T_1(0,Q^2)	\nonumber\\
	& + \frac{q_0^2}{2\pi M} \int_{\nu_{th}}^\infty  d\nu 
	\frac{F_1(\nu,Q^2)}{\nu (\nu^2 - q_0^2) }	\,.
\end{align}
The nucleon pole is isolated in $T_1^{pole}$ and
the integral begins at the inelastic threshold $\nu_{th} = (2Mm_\pi + m_\pi^2 + Q^2)/(2M)$.  Similarly, as $T_2^B$ contains only a pole term,
\begin{align}
T_2(q_0,Q^2) &= T_2^B(q_0,Q^2)  + \frac{1}{2\pi} \int_{\nu_{th}}^\infty  d\nu 
	\frac{ F_2(\nu,Q^2) }{ \nu^2 - q_0^2 }	\,.
\end{align}

With
\be
\Delta E = \Delta E^{subt} + \Delta E^{inel} + \Delta E^{el}		\,,
\ee
we obtain
\begin{align}
&\Delta E^{subt} =  \frac{4\pi\alpha^2}{m} \phi^2_n(0) \int_0^\infty \frac{dQ^2}{Q^2}
	 \frac{\gamma_1(\tau_\ell)}{\sqrt{\tau_\ell}}  \overline T_1(0,Q^2)	\,,
					\end{align}\begin{align}
\label{eq:inel}
&\Delta E^{inel} = - \frac{2\alpha^2}{mM} \phi^2_n(0) \int_0^\infty \frac{dQ^2}{Q^2}
					\nonumber\\
&\ 	\times	\int_{\nu_{th}}^\infty d\nu 
	\left[	\frac{ \widetilde\gamma_1(\tau,\tau_\ell) F_1(\nu,Q^2)}{\nu}
	+	\frac{ \widetilde\gamma_2(\tau,\tau_\ell) F_2(\nu,Q^2)}{Q^2/M}
	\right]	,
					\\
&\Delta E^{el} = -\frac{\alpha^2 m}{M(M^2-m^2)}  \phi^2_n(0)
	\int_0^\infty \frac{dQ^2}{Q^2}  	\nonumber \\
&\qquad \times	\bigg\{
	\left( \frac{ \gamma_2(\tau_p)}{\sqrt{\tau_p}} 
	- \frac{ \gamma_2(\tau_\ell)}{\sqrt{\tau_\ell}}	\right)
	\frac{ G_E^2 + \tau_p G_M^2 } { \tau_p (1+\tau_p) }
							\nonumber\\
&\qquad\ \ 	- \left( \frac{ \gamma_1(\tau_p)}{\sqrt{\tau_p}} 
	- \frac{ \gamma_1(\tau_\ell)}{\sqrt{\tau_\ell}}	\right)  G_M^2
	\bigg\},
\end{align}
where $\tau = \nu^2/Q^2$ and $\tau_\ell = Q^2/(4m^2)$.  The auxiliary functions are
\begin{align}
\gamma_1(\tau) &= (1-2\tau) \Big((1+\tau)^{1/2} - \tau^{1/2} \Big) + \tau^{1/2}	,
\nonumber\\
\gamma_2(\tau) &= (1+\tau)^{3/2} - \tau^{3/2} - \frac{3}{2}\tau^{1/2}	.
\end{align}
Both are monotonically falling functions, reducing to 1 at $\tau=0$ and falling like $\tau^{-1/2}$ at large $\tau$.  Also
\begin{align}
\widetilde\gamma_1(\tau,\tau_\ell) &= \frac{1}{\tau_\ell - \tau}\Big(
	\sqrt{\tau_\ell} \gamma_1(\tau_\ell) - \sqrt{\tau} \gamma_1(\tau)      \Big)	,
\nonumber\\
\widetilde\gamma_2(\tau,\tau_\ell) &= \frac{1}{\tau_\ell - \tau}\left(\frac{ \gamma_2(\tau)}	{\sqrt{\tau}} - \frac{\gamma_2(\tau_\ell)}{\sqrt{\tau_\ell} }		\right)  .
\end{align}





The subtraction function $\overline T(0,Q^2)$ has unphysical arguments, excepting the point $Q^2 = 0$.  It comes from the excitation of the proton, and can at low $Q^2$ (and low $\nu$, in general) be described using the electric ($\alpha_E$) and magnetic ($\beta_M$) polarizabilities and the effective Hamiltonian
\be
\mathcal H = - \frac{1}{2} 4\pi \alpha_E \vec E^{\,2} 
	- \frac{1}{2} 4\pi \beta_M \vec B^{\,2}	.
\ee
For small $\nu$ and $Q$, this gives
\be
\lim_{\nu^2,Q^2\to 0}  \overline T_1(\nu,Q^2) =  \frac{\nu^2}{e^2} \left( \alpha_E +\beta_M \right)
	+ \frac{Q^2}{e^2} \beta_M	\,.
\ee
The $\nu^2$ term is shown to connect to known results in another context~\cite{Drechsel:2002ar}, and the $Q^2$ term was obtained by Pachucki~\cite{Pachucki:1996zza}.   With the above result, the integral over $\overline T_1(0,Q^2)$ converges at the lower limit. 

\begin{figure}[htbp]
\includegraphics[width = 3.4 in]{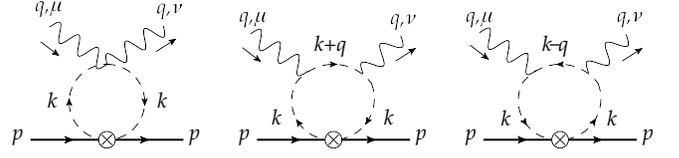}
\caption{Diagrams used for estimating the $Q^2$ dependence of the non-pole part of the subtraction term.}
\label{fig:ggsigma}
\end{figure}

For higher $Q^2$, the subtraction function comes from non-nucleon-pole contributions, and the forward amplitude is dominated by low mass intermediate states.  With the $Q^2 \to 0$ limit fixed in terms of $\beta_M$, we estimate the $Q^2$ dependence from pion loop contributions where the two-pion state has a scalar coupling to the nucleon, as illustrated in Fig.~\ref{fig:ggsigma}.

With standard Feynman rules for scalar QED and an effective $g_S\bar N \phi^\dagger \phi N$ coupling for the lower vertex, one obtains from these terms
\ba
\overline T^{\mu\nu}= \frac{ g_S }{192 \pi^3 m_\pi^2}
	\left( q^2 g^{\mu\nu} - q^\mu q^\nu \right)		F_{loop}(Q^2)	\,,
\ea
where, with $\lambda = 4m_\pi^2/Q^2$,
\begin{align}
&F_{loop}(Q^2) = \frac{3\lambda}{2} 
	\left\{ \sqrt{1+ \lambda}
	\ln\frac{\sqrt{1+\lambda}+1}{\sqrt{1+ \lambda}-1} - 2 \right\}
					\nonumber\\
& \quad = \left\{	\begin{array}{ll}
	1 - \frac{Q^2}{10m_\pi^2} + \mathcal O(\frac{Q^4}{m_\pi^2})
		 \,, & Q \to 0 	\\[1ex]
	\frac{6m_\pi^2}{Q^2} \left( \ln\frac{Q^2}{m_\pi^2} - 2 \right) + 
		O(\frac{m_\pi^4}{Q^4}) \,,
			& Q \to \infty
	\end{array}	\right.	\,.
\end{align}
We can identify
$
\beta_M = {\alpha g_S}/{48\pi^2 m_\pi^2}		\,,
$
and obtain
\be
\label{eq:t1}
\overline T_1(0,Q^2) = \frac{\beta_M}{4\pi\alpha} Q^2 F_{loop}(Q^2)	\,.
\ee




The Particle Data Group gives~\cite{Nakamura:2010zzi},
\be
\beta_M =  (1.9 \pm 0.5) \times 10^{-4} {\rm\ fm}^3 	.
\ee
However, according to some recent analyses,
\be
\beta_M = \left\{
\begin{array}{cl}
(4.0\pm 0.7) \times 10^{-4} {\rm\ fm}^3
& \rm{ [18]}
\\
(3.4\pm 1.2)\times 10^{-4} {\rm\ fm}^3
& \rm{ [19,20] \,. }
\end{array}
\right.
\ee

Using the subtraction function from Eq.~(\ref{eq:t1}), we find
\be
\Delta E^{subt} = 5.3 \ \mu{\rm eV} \times 
	\frac{\beta_M}{ (3.4 \times 10^{-4} {\rm\, fm^3}) }	\,.
\ee
Much of the support for the integral is at low $Q^2$, being controlled by 
$\gamma_1$ as well as by the $Q^2$ dependence from the pion loop, and half the contributions to $\Delta E^{subt}$ come from $Q^2 \lesssim 0.04$ GeV$^2$, albeit with a long tail.

Refs.~\cite{Pachucki:1999zza} and~\cite{Martynenko:2005rc} found $\Delta E^{subt}$ to be $1.8$ and $2.3$ $\mu$eV, using $\beta_M = 1.5$ and $1.9 \times 10^{-4}$ fm$^3$, respectively, and using a $Q^2$ falloff related to the nucleon electromagnetic form factor.  For the same $\beta_M$, our results are about 30\% larger due to having flatter $Q^2$ falloff.  

One can also consider inserting a form factor $F_\pi$ for each incoming photon coupling to pions, modifying the subtraction function of Eq.~(\ref{eq:t1}) by multiplying it with  $F_\pi(Q^2)^2$.
Obtaining $F_\pi$ from the fit of~\cite{Huber:2008id}, we find $\Delta E^{subt} = 3.8\ \mu$eV.

The inelastic contributions depend on $F_{(1,2)}(\nu,Q^2)$, and good data in the low-$Q^2$ and resonance region is available from Jefferson Lab. Analytic representations of this data are given by Christy and Bosted~\cite{Christy:2007ve}, in a fit valid for $0<Q^2<8$ GeV$^2$ and $W$ from threshold to $3.1$ GeV, where $W$ is the final hadronic mass for inelastic $ep$ scattering, $W^2 = M^2 + 2M\nu -Q^2$.   From the Bosted-Christy region, we obtain a 
$-12.2 \ \mu$eV contribution to $\Delta E^{inel}$.  We also use the fit of Capella \textit{et al.}~\cite{Capella:1994cr}, valid for data at low and intermediate $Q^2$ above the resonance region, specifically $0<Q^2<5$ GeV$^2$ and $W > 2.5$ GeV.  This gives a $-0.5 \ \mu$eV contribution using~\cite{Capella:1994cr} for $W>3.1$ GeV in the allowed $Q^2$ region.  Contributions from higher $Q^2$ are quite small (on the order of 0.002 $\mu$eV from $Q>5$ GeV$^2$ and $W>3.1$ GeV).  We thus have
\be
\Delta E^{inel} = -12.7 \ \mu{\rm eV}.
\ee
Refs.~\cite{Pachucki:1999zza} and~\cite{Martynenko:2005rc} quoted $-13.9$ and $-13.8$ $\mu$eV for this contribution.

The elastic contribution depends on the nucleon form factors, for a selection of modern form factors we get
\be
\Delta E^{el} = \left\{		\begin{array}{cl}
	- 27.8 \ \mu{\rm eV}	& {\rm Kelly~[25]	 }		\\
	- 29.5 \ \mu{\rm eV}	& {\rm AMT~[26]	 }		\\
	- 30.8 \ \mu{\rm eV}	& 
		{\rm Mainz~2010~[4,27] }
	\end{array}	\,.
	\right.
\ee
Ref.~\cite{Pachucki:1996zza} quoted $-23 \ \mu$eV using the Simon \textit{et al.}~form factors~\cite{Simon:1980hu} from 1980.  However, the main difference between our results is not due to the newness of the form factors, but rather to our exclusion of the non-pole contributions from the elastic contributions.  The non-pole contributions would be a positive $4.7$ (Kelly) or $4.8 \ \mu$eV (AMT or Mainz 2010) contribution were they included.




Table~\ref{table:one} summarizes our numerical results (selecting the AMT form factors~\cite{Arrington:2007ux} for the elastic terms)  and compares them to earlier work.  (Ref.~\cite{Martynenko:2005rc} did not calculate the elastic term, so we carried over the result from~\cite{Pachucki:1996zza}.)  

\begin{table}[htdp]
\caption{Numerical results for the $\mathcal O(\alpha^5)$ proton structure corrections to the Lamb shift in muonic hydrogen.  Energies are in $\mu$eV.}
\begin{center}
\begin{ruledtabular}
\begin{tabular}{lccc}
($\mu$eV)	&	this work &  Ref.~\cite{Pachucki:1996zza,Pachucki:1999zza}  
			& Ref.~\cite{Martynenko:2005rc} \\
			\hline
$\Delta E^{subt}$  & $\quad\ \, 5.3 \pm 1.9$ & $\quad\ \, 1.8$ & $\quad\ \, 2.3$ \\
$\Delta E^{inel}$  & $-12.7 \pm 0.5$  & $-13.9$ & $-13.8$ \\
$\Delta E^{el}$  & $-29.5 \pm 1.3$ & $-23.0$ & $-23.0$ \\
\hline
$\Delta E$  & $-36.9 \pm 2.4$ & $-35.1$ & $-34.5$ \\
\end{tabular}
\end{ruledtabular}
\end{center}
\label{table:one}
\end{table}%

Regarding the uncertainties, for the subtraction term energy, we took  $\beta_M$ from~\cite{Beane:2004ra}, and propagated their error limits, which are large enough to accommodate the other two $\beta_M$ values.  The inelastic energy, comes mainly from~\cite{Christy:2007ve}, which states that most of the data points are fit to within 3\%. The data itself typically had 3\% error limits, and we added these two errors in quadrature.  For the elastic term, we estimated the error from the spread between the two newer form factor fits that we used.  We added the errors in quadrature to obtain the total error.

Our results are similar to previous results in aggregate.  This seems to be happenstance, since changes in the individual contributions are larger than the change in the total.  The main changes occurred because we feel use of a larger magnetic polarizability is justified and because using a dispersive treatment throughout does not allow keeping the elastic non-pole contributions.


\begin{acknowledgments}

CEC thanks the National Science Foundation for support under Grant PHY-0855618 and thanks the Helmholtz Gemeinschaft in Mainz and the Helsinki Institute for Physics for their hospitality.  We thank Vladimir Pascalutsa for useful comments.

\end{acknowledgments}


\bibliography{MuonicLamb}

\end{document}